

%
%


\def\famname{
 \textfont0=\textrm \scriptfont0=\scriptrm
 \scriptscriptfont0=\sscriptrm
 \textfont1=\textmi \scriptfont1=\scriptmi
 \scriptscriptfont1=\sscriptmi
 \textfont2=\textsy \scriptfont2=\scriptsy \scriptscriptfont2=\sscriptsy
 \textfont3=\textex \scriptfont3=\textex \scriptscriptfont3=\textex
 \textfont4=\textbf \scriptfont4=\scriptbf \scriptscriptfont4=\sscriptbf
 \skewchar\textmi='177 \skewchar\scriptmi='177
 \skewchar\sscriptmi='177
 \skewchar\textsy='60 \skewchar\scriptsy='60
 \skewchar\sscriptsy='60
 \def\rm{\fam0 \textrm} \def\bf{\fam4 \textbf}}
\def\sca#1{scaled\magstep#1} \def\scah{scaled\magstephalf} 
\def\twelvepoint{
 \font\textrm=cmr12 \font\scriptrm=cmr8 \font\sscriptrm=cmr6
 \font\textmi=cmmi12 \font\scriptmi=cmmi8 \font\sscriptmi=cmmi6 
 \font\textsy=cmsy10 \sca1 \font\scriptsy=cmsy8
 \font\sscriptsy=cmsy6
 \font\textex=cmex10 \sca1
 \font\textbf=cmbx12 \font\scriptbf=cmbx8 \font\sscriptbf=cmbx6
 \font\it=cmti12
 \font\sectfont=cmbx12 \sca1
 \font\sectmath=cmmib10 \sca2
 \font\sectsymb=cmbsy10 \sca2
 \font\refrm=cmr10 \scah \font\refit=cmti10 \scah
 \font\refbf=cmbx10 \scah
 \def\twelverm{\textrm} \def\twelveit{\it} \def\twelvebf{\textbf}
 \famname \textrm 
 \advance\voffset by .06in \advance\hoffset by .28in
 \normalbaselineskip=17.5pt plus 1pt \baselineskip=\normalbaselineskip
 \parindent=21pt
 \setbox\strutbox=\hbox{\vrule height10.5pt depth4pt width0pt}}


\catcode`@=11

{\catcode`\'=\active \def'{{}^\bgroup\prim@s}}

\def\screwcount{\alloc@0\count\countdef\insc@unt}   
\def\screwdimen{\alloc@1\dimen\dimendef\insc@unt} 
\def\screwbox{\alloc@4\box\chardef\insc@unt}

\catcode`@=12


\overfullrule=0pt			
\vsize=9in \hsize=6in
\lineskip=0pt				
\abovedisplayskip=1.2em plus.3em minus.9em 
\belowdisplayskip=1.2em plus.3em minus.9em	
\abovedisplayshortskip=0em plus.3em	
\belowdisplayshortskip=.7em plus.3em minus.4em	
\parindent=21pt
\setbox\strutbox=\hbox{\vrule height10.5pt depth4pt width0pt}
\def\makefootline{\baselineskip=30pt \line{\the\footline}}
\footline={\ifnum\count0=1 \hfil \else\hss\twelverm\folio\hss \fi}
\pageno=1


\def\put(#1,#2)#3{\screwdimen\unit  \unit=1in
	\vbox to0pt{\kern-#2\unit\hbox{\kern#1\unit
	\vbox{#3}}\vss}\nointerlineskip}


\def\\{\hfil\break}
\def\newpage{\vfill\eject}
\def\center{\leftskip=0pt plus 1fill \rightskip=\leftskip \parindent=0pt
 \def\textindent##1{\par\hangindent21pt\footrm\noindent\hskip21pt
 \llap{##1\enspace}\ignorespaces}\par}
\def\unnarrower{\leftskip=0pt \rightskip=\leftskip}


\def\vol#1 {{\refbf#1} }		 


\def\NP #1 {{\refit Nucl. Phys.} {\refbf B{#1}} }
\def\PL #1 {{\refit Phys. Lett.} {\refbf{#1}} }
\def\PR #1 {{\refit Phys. Rev. Lett.} {\refbf{#1}} }
\def\PRD #1 {{\refit Phys. Rev.} {\refbf D{#1}} }


\hyphenation{pre-print}
\hyphenation{quan-ti-za-tion}

%
%


\def\oonoo#1#2#3{\vbox{\ialign{##\crcr
	\hfil\hfil\hfil{$#3{#1}$}\hfil\crcr\noalign{\kern1pt\nointerlineskip}
	$#3{#2}$\crcr}}}
\def\oon#1#2{\mathchoice{\oonoo{#1}{#2}{\displaystyle}}
	{\oonoo{#1}{#2}{\textstyle}}{\oonoo{#1}{#2}{\scriptstyle}}
	{\oonoo{#1}{#2}{\scriptscriptstyle}}}
\def\dt#1{\oon{\hbox{\bf .}}{#1}}  
\def\ddt#1{\oon{\hbox{\bf .\kern-1pt.}}#1}    
\def\slap#1#2{\setbox0=\hbox{$#1{#2}$}
	#2\kern-\wd0{\hfuzz=1pt\hbox to\wd0{\hfil$#1{/}$\hfil}}}
\def\sla#1{\mathpalette\slap{#1}}                
\def\bop#1{\setbox0=\hbox{$#1M$}\mkern1.5mu
	\lower.02\ht0\vbox{\hrule height0pt depth.06\ht0
	\hbox{\vrule width.06\ht0 height.9\ht0 \kern.9\ht0
	\vrule width.06\ht0}\hrule height.06\ht0}\mkern1.5mu}
\def\bo{{\mathpalette\bop{}}}                        
\def~{\widetilde} 
\mathcode`\*="702A                  
\def\in{\relax\ifmmode\mathchar"3232\else{\refit in\/}\fi} 
\def\half{{\textstyle{1\over{\raise.1ex\hbox{$\scriptstyle{2}$}}}}}

\def\Gamma{\mathchar"0100}
\def\Delta{\mathchar"0101}
\def\Theta{\mathchar"0102}
\def\Lambda{\mathchar"0103}
\def\Xi{\mathchar"0104}
\def\Pi{\mathchar"0105}
\def\Sigma{\mathchar"0106}
\def\Upsilon{\mathchar"0107}
\def\Phi{\mathchar"0108}
\def\Psi{\mathchar"0109}
\def\Omega{\mathchar"010A}

\catcode128=13 \def €{\"A}                 
\catcode129=13 \def {\AA}                 
\catcode130=13 \def '{\c}           	   
\catcode131=13 \def ƒ{\'E}                   
\catcode132=13 \def "{\~N}                   
\catcode133=13 \def …{\"O}                 
\catcode134=13 \def †{\"U}                  
\catcode135=13 \def ‡{\'a}                  
\catcode136=13 \def ˆ{\`a}                   
\catcode137=13 \def ‰{\^a}                 
\catcode138=13 \def Š{\"a}                 
\catcode139=13 \def ‹{\~a}                   
\catcode140=13 \def Œ{\alpha}            
\catcode141=13 \def {\chi}                
\catcode142=13 \def Ž{\'e}                   
\catcode143=13 \def {\`e}                    
\catcode144=13 \def {\^e}                  
\catcode145=13 \def '{\"e}                
\catcode146=13 \def '{\'\i}                 
\catcode147=13 \def "{\`\i}                  
\catcode148=13 \def "{\^\i}                
\catcode149=13 \def •{\"\i}                
\catcode150=13 \def –{\~n}                  
\catcode151=13 \def —{\'o}                 
\catcode152=13 \def ˜{\`o}                  
\catcode153=13 \def ™{\^o}                
\catcode154=13 \def š{\"o}                 
\catcode155=13 \def ›{\~o}                  
\catcode156=13 \def œ{\'u}                  
\catcode157=13 \def {\`u}                  
\catcode158=13 \def ž{\^u}                
\catcode159=13 \def Ÿ{\"u}                
\catcode160=13 \def  {\tau}               
\catcode161=13 \mathchardef ¡="2203     
\catcode162=13 \def ¢{\oplus}           
\catcode163=13 \def £{\relax\ifmmode\to\else\itemize\fi} 
\catcode164=13 \def ¤{\subset}	  
\catcode165=13 \def ¥{\infty}           
\catcode166=13 \def ¦{\mp}                
\catcode167=13 \def §{\sigma}           
\catcode168=13 \def ¨{\rho}               
\catcode169=13 \def ©{\gamma}         
\catcode170=13 \def ª{\leftrightarrow} 
\catcode171=13 \def «{\relax\ifmmode\acute\else\expandafter\'\fi}
\catcode172=13 \def ¬{\relax\ifmmode\expandafter\ddt\else\expandafter\"\fi}
\catcode173=13 \def ­{\equiv}            
\catcode174=13 \def ®{\approx}          
\catcode175=13 \def ¯{\Omega}          
\catcode176=13 \def °{\otimes}          
\catcode177=13 \def ±{\ne}                 
\catcode178=13 \def ²{\le}                   
\catcode179=13 \def ³{\ge}                  
\catcode180=13 \def ´{\upsilon}          
\catcode181=13 \def µ{\mu}                
\catcode182=13 \def ¶{\delta}             
\catcode183=13 \def ·{\epsilon}          
\catcode184=13 \def ¸{\Pi}                  
\catcode185=13 \def ¹{\pi}                  
\catcode186=13 \def º{\beta}               
\catcode187=13 \def »{\partial}           
\catcode188=13 \def ¼{\nobreak\ }       
\catcode189=13 \def ½{\zeta}               
\catcode190=13 \def ¾{\sim}                 
\catcode191=13 \def ¿{\omega}           
\catcode192=13 \def À{\dt}                     
\catcode193=13 \def Á{\gets}                
\catcode194=13 \def Â{\lambda}           
\catcode195=13 \def Ã{\nu}                   
\catcode196=13 \def Ä{\phi}                  
\catcode197=13 \def Å{\xi}                     
\catcode198=13 \def Æ{\psi}                  
\catcode199=13 \def Ç{\int}                    
\catcode200=13 \def È{\oint}                 
\catcode201=13 \def É{\relax\ifmmode\cdot\else\vol\fi}    
\catcode202=13 \def Ê{\relax\ifmmode\,\else\thinspace\fi}
\catcode203=13 \def Ë{\`A}                      
\catcode204=13 \def Ì{\~A}                      
\catcode205=13 \def Í{\~O}                      
\catcode206=13 \def Î{\Theta}              
\catcode207=13 \def Ï{\theta}               
\catcode208=13 \def Ð{\relax\ifmmode\bar\else\expandafter\=\fi}
\catcode209=13 \def Ñ{\overline}             
\catcode210=13 \def Ò{\langle}               
\catcode211=13 \def Ó{\relax\ifmmode\{\else\ital\fi}      
\catcode212=13 \def Ô{\rangle}               
\catcode213=13 \def Õ{\}}                        
\catcode214=13 \def Ö{\sla}                      
\catcode215=13 \def ×{\relax\ifmmode\check\else\expandafter\v\fi}
\catcode216=13 \def Ø{\"y}                     
\catcode217=13 \def Ù{\"Y}  		    
\catcode218=13 \def Ú{\Leftarrow}       
\catcode219=13 \def Û{\Leftrightarrow}       
\catcode220=13 \def Ü{\relax\ifmmode\Rightarrow\else\sect\fi}
\catcode221=13 \def Ý{\sum}                  
\catcode222=13 \def Þ{\prod}                 
\catcode223=13 \def ß{\widehat}              
\catcode224=13 \def à{\pm}                     
\catcode225=13 \def á{\nabla}                
\catcode226=13 \def â{\quad}                 
\catcode227=13 \def ã{\in}               	
\catcode228=13 \def ä{\star}      	      
\catcode229=13 \def å{\sqrt}                   
\catcode230=13 \def æ{\^E}			
\catcode231=13 \def ç{\Upsilon}              
\catcode232=13 \def è{\"E}    	   	 
\catcode233=13 \def é{\`E}               	  
\catcode234=13 \def ê{\Sigma}                
\catcode235=13 \def ë{\Delta}                 
\catcode236=13 \def ì{\Phi}                     
\catcode237=13 \def í{\`I}        		   
\catcode238=13 \def î{\iota}        	     
\catcode239=13 \def ï{\Psi}                     
\catcode240=13 \def ð{\times}                  
\catcode241=13 \def ñ{\Lambda}             
\catcode242=13 \def ò{\cdots}                
\catcode243=13 \def ó{\^U}			
\catcode244=13 \def ô{\`U}    	              
\catcode245=13 \def õ{\bo}                       
\catcode246=13 \def ö{\relax\ifmmode\hat\else\expandafter\^\fi}
\catcode247=13 \def÷{\relax\ifmmode\tilde\else\expandafter\~\fi}
\catcode248=13 \def ø{\ll}                         
\catcode249=13 \def ù{\gg}                       
\catcode250=13 \def ú{\eta}                      
\catcode251=13 \def û{\kappa}                  
\catcode252=13 \def ü{\half}     		 
\catcode253=13 \def ý{\Gamma} 		
\catcode254=13 \def þ{\Xi}   			
\catcode255=13 \def ÿ{\relax\ifmmode{}^{\dagger}{}\else\dag\fi}


\def\ital#1Õ{{\it#1\/}}	     
\def\un#1{\relax\ifmmode\underline#1\else $\underline{\hbox{#1}}$
	\relax\fi}

\def\roonoo#1#2#3{\vbox{\ialign{##\crcr
	\hfil{$#3{#1}$}\hfil\crcr\noalign{\kern1pt\nointerlineskip}
	$#3{#2}$\crcr}}}

\def\tdt#1{\oon{\hbox{\bf .\kern-1pt.\kern-1pt.}}#1}   
\def\({\eqno(}



\def\õ#1{
	\screwcount\num
	\num=1
	\screwdimen\downsy
	\downsy=-1.5ex
	\mkern-3.5mu
	õ
	\loop
	\ifnum\num<#1
	\llap{\raise\num\downsy\hbox{$õ$}}
	\advance\num by1
	\repeat}
\def\upõ#1#2{\screwcount\numup
	\numup=#1
	\advance\numup by-1
	\screwdimen\upsy
	\upsy=.75ex
	\mkern3.5mu
	\raise\numup\upsy\hbox{$#2$}}



\newcount\marknumber	\marknumber=1
\newcount\countdp \newcount\countwd \newcount\countht 

%
%
\ifx\pdfoutput\undefined
\def\rgboo#1{}
\input epsf

\def\postscript#1{\special{" #1}}		
\postscript{
	/bd {bind def} bind def
	/fsd {findfont exch scalefont def} bd
	/sms {setfont moveto show} bd
	/ms {moveto show} bd
	/pdfmark where		
	{pop} {userdict /pdfmark /cleartomark load put} ifelse
	[ /PageMode /UseOutlines		
	/DOCVIEW pdfmark}
\def\bookmark#1#2{\postscript{		
	[ /Dest /MyDest\the\marknumber /View [ /XYZ null null null ] /DEST pdfmark
	[ /Title (#2) /Count #1 /Dest /MyDest\the\marknumber /OUT pdfmark}%
	\advance\marknumber by1}
\def\pdfklink#1#2{%
	\hskip-.25em\setbox0=\hbox{#1}%
		\countdp=\dp0 \countwd=\wd0 \countht=\ht0%
		\divide\countdp by65536 \divide\countwd by65536%
			\divide\countht by65536%
		\advance\countdp by1 \advance\countwd by1%
			\advance\countht by1%
		\def\linkdp{\the\countdp} \def\linkwd{\the\countwd}%
			\def\linkht{\the\countht}%
	\postscript{
		[ /Rect [ -1.5 -\linkdp.0 0\linkwd.0 0\linkht.5 ] 
		/Border [ 0 0 0 ]
		/Action << /Subtype /URI /URI (#2) >>
		/Subtype /Link
		/ANN pdfmark}{\rgb{1 0 0}{#1}}}
%
%
\else
\def\rgboo#1{\pdfliteral{#1 rg #1 RG}}

\pdfcatalog{/PageMode /UseOutlines}		
\def\bookmark#1#2{
	\pdfdest num \marknumber xyz
	\pdfoutline goto num \marknumber count #1 {#2}
	\advance\marknumber by1}
\def\pdfklink#1#2{%
	\noindent\pdfstartlink user
		{/Subtype /Link
		/Border [ 0 0 0 ]
		/A << /S /URI /URI (#2) >>}{\rgb{1 0 0}{#1}}%
	\pdfendlink}
\fi

\def\rgbo#1#2{\rgboo{#1}#2\rgboo{0 0 0}}
\def\rgb#1#2{\mark{#1}\rgbo{#1}{#2}\mark{0 0 0}}
\def\pdflink#1{\pdfklink{#1}{#1}}
\def\xxxlink#1{\pdfklink{[arXiv:#1]}{http://arXiv.org/abs/#1}}

\catcode`@=11

\def\wlog#1{}	


\def\makeheadline{\vbox to\z@{\vskip-36.5\p@
	\line{\vbox to8.5\p@{}\the\headline%
	\ifnum\pageno=\z@\rgboo{0 0 0}\else\rgboo{\topmark}\fi%
	}\vss}\nointerlineskip}
\headline={
	\ifnum\pageno=\z@
		\hfil
	\else
		\ifnum\pageno<\z@
			\ifodd\pageno
				\tenrm\romannumeral-\pageno\hfil\lefthead\hfil
			\else
				\tenrm\hfil\righthead\hfil\romannumeral-\pageno
			\fi
		\else
			\ifodd\pageno
				\tenrm\hfil\righthead\hfil\number\pageno
			\else
				\tenrm\number\pageno\hfil\lefthead\hfil
			\fi
		\fi
	\fi}

\catcode`@=12

\def\righthead{\hfil} \def\lefthead{\hfil}
\nopagenumbers


\def\chrulefill{\rgb{1 0 0}{\hrulefill}}
\def\cdotfill{\rgb{1 0 0}{\dotfill}}
\newcount\area	\area=1
\newcount\cross	\cross=1
\def\volume#1\par{\newpage\noindent{\biggest{\rgb{1 .5 0}{#1}}}
	\par\nobreak\bigskip\medskip\area=0}
\def\chapskip{\par\ifnum\area=0\bigskip\medskip\goodbreak
	\else\newpage\fi}
\def\chapy#1{\area=1\cross=0
	\xdef\lefthead{\rgbo{1 0 .5}{#1}}\vbox{\biggerer\offinterlineskip
	\line{\chrulefill¼\hphantom{\lefthead}\chrulefill}
	\line{\chrulefill¼\lefthead\chrulefill}}\par\nobreak\medskip}
\def\chap#1\par{\chapskip\bookmark3{#1}\chapy{#1}}
\def\sectskip{\par\ifnum\cross=0\bigskip\medskip\goodbreak
	\else\newpage\fi}
\def\secty#1{\cross=1
	\xdef\righthead{\rgbo{1 0 1}{#1}}\vbox{\bigger\offinterlineskip
	\line{\cdotfill¼\hphantom{\righthead}\cdotfill}
	\line{\cdotfill¼\righthead\cdotfill}}\par\nobreak\medskip}
\def\sect#1 #2\par{\sectskip\bookmark{#1}{#2}\secty{#2}}
\def\subsectskip{\par\ifdim\lastskip<\medskipamount
	\bigskip\medskip\goodbreak\else\nobreak\fi}
\def\subsecty#1{\noindent{\sectfont{\rgbo{.5 0 1}{#1}}}\par\nobreak\medskip}
\def\subsect#1\par{\subsectskip\bookmark0{#1}\subsecty{#1}}
\long\def\x#1 #2\par{\hangindent2\parindent%
\mark{0 0 1}\rgboo{0 0 1}{\bf Exercise #1}\\#2%
\par\rgboo{0 0 0}\mark{0 0 0}}
\def\refs{\bigskip\noindent{\bf \rgbo{0 .5 1}{REFERENCES}}\par\nobreak\medskip
	\frenchspacing \parskip=0pt \refrm \baselineskip=1.23em plus 1pt
	\def\ital##1Õ{{\refit##1\/}}}
\long\def\twocolumn#1#2{\hbox to\hsize{\vtop{\hsize=2.9in#1}
	\hfil\vtop{\hsize=2.9in #2}}}


\twelvepoint
\font\bigger=cmbx12 \sca2
\font\biggerer=cmb10 \sca5
\font\biggest=cmssdc10 scaled 2850
 \sca5

 \sca3


\def Ü{\relax\ifmmode\Rightarrow\else\expandafter\subsect\fi}
\def Û{\relax\ifmmode\Leftrightarrow\else\expandafter\sect\fi}
\def Ú{\relax\ifmmode\Leftarrow\else\expandafter\chap\fi}

\def\itemize#1 {\item{\bf#1}}
\def\itemizze#1 {\itemitem{\bf#1}}
\def\itemutem{\par\indent\indent \hangindent3\parindent \textindent}
\def\itemizzze#1 {\itemutem{\bf#1}}
\def ª{\relax\ifmmode\leftrightarrow\else\itemizze\fi}
\def Á{\relax\ifmmode\gets\else\itemizzze\fi}

\def\¢{\ominus}

\def\Ä{\varphi}  \def\¿{\varpi}	\def\Ï{\vartheta}

\def ò{\relax\ifmmode\cdots\else\dotfill\fi}

\chardef\slo="1C


\def\cvrule{\rgbo{0 .5 1}{\vrule}}
\def\chrule{\rgbo{0 .5 1}{\hrule}}
\def\boxit#1{\leavevmode\thinspace\hbox{\cvrule\vtop{\vbox{\chrule%
	\vskip3pt\kern1pt\hbox{\vphantom{\bf/}\thinspace\thinspace%
	{\bf#1}\thinspace\thinspace}}\kern1pt\vskip3pt\chrule}\cvrule}%
	\thinspace}
\def\Boxit#1{\noindent\vbox{\chrule\hbox{\cvrule\kern3pt\vbox{
	\advance\hsize-7pt\vskip-\parskip\kern3pt\bf#1
	\hbox{\vrule height0pt depth\dp\strutbox width0pt}
	\kern3pt}\kern3pt\cvrule}\chrule}}




\def\today{\ifcase\month\or
 January\or February\or March\or April\or May\or June\or July\or
 August\or September\or October\or November\or December\fi
 \space\number\day, \number\year}

\parindent=20pt
\newskip\normalparskip	\normalparskip=.7\medskipamount
\parskip=\normalparskip	



\catcode`\|=\active \catcode`\<=\active \catcode`\>=\active 
\def|{\relax\ifmmode\delimiter"026A30C \else$\mathchar"026A$\fi}
\def<{\relax\ifmmode\mathchar"313C \else$\mathchar"313C$\fi}
\def>{\relax\ifmmode\mathchar"313E \else$\mathchar"313E$\fi}


%
%
%
%
%
%
%

\def\thetitle#1#2#3#4#5{
 \def\titlefont{\biggest} \font\footrm=cmr10 \font\footit=cmti10
  \twelverm
	{\hbox to\hsize{#4 \hfill YITP-SB-#3}}\par
	\vskip.8in minus.1in {\center\baselineskip=2.2\normalbaselineskip
 {\titlefont #1}\par}{\center\baselineskip=\normalbaselineskip
 \vskip.5in minus.2in #2
	\vskip1.4in minus1.2in {\twelvebf ABSTRACT}\par}
 \vskip.1in\par
 \narrower\par#5\par\unnarrower\vskip3.5in minus3.3in\eject}
\def\paper\par#1\par#2\par#3\par#4\par#5\par{
	\thetitle{#1}{#2}{#3}{#4}{#5}} 
\def\author#1#2{#1 \vskip.1in {\twelveit #2}\vskip.1in}
\def\YITP{C. N. Yang Institute for Theoretical Physics\\
	State University of New York, Stony Brook, NY 11794-3840}
\def\WS{W. Siegel\footnote{$*$}{
	\pdflink{mailto:siegel@insti.physics.sunysb.edu}\\
	\pdfklink{http://insti.physics.sunysb.edu/\~{}siegel/plan.html}
	{http://insti.physics.sunysb.edu/\noexpand~siegel/plan.html}}}


\pageno=0

\paper

\vskip-.5in
{\rgb{1 .3 0}{F-theory with zeroth-quantized ghosts}}

\author\WS\YITP
\vskip-.1in

16-2

January 15, 2016

F-theory in its most general sense should be a theory defined on a worldvolume of higher dimension than the worldsheet, that reproduces string results perturbatively but includes nonperturbative supergravity solutions at the first-quantized level.  This implies that in some sense it should contain the same oscillator modes as the string but an enlarged set of zero-modes.  In this paper we concentrate on the higher-dimensional properties of the worldvolume (rather than those of spacetime):  ``Ghost" dimensions are added to the worldvolume, as might be expected in a ``zeroth-quantized" approach to the constraints on its higher bosonic dimensions, by adding equal numbers of bosonic and fermionic dimensions to the worldsheet.

\pageno=2

Ü1. Introduction

F-theory was originally introduced as a method to find compactified vacua more general than those found from supergravities in D=10 and 11 [1].  ``Exceptional supergravities" [2] were defined to allow this procedure without incorporating string excitations (but have not yet reached completion for the most symmetric cases).  However, these excitations are necessary for the improved high-energy behavior that made string theory an attractive approach to quantum gravity in the first place.  Then, as M(embrane)-theory [3] was an enlargement of the string theory worldvolume from d=2 to 3 in association with the increase of spacetime dimension from D=10 to 11, F-theory would be a further (set of) step(s) to increase both.  Unfortunately, it seems that quantizing the (super)membrane in the obvious way did not include supergravity among its states [4].

This led us to consider a new approach to the worldvolume [5], where its extra dimensions were constrained in the same way as those of spacetime in manifestly T-dual string theory [6] or exceptional supergravities.  Although these F-theories have been shown equivalent to string theory, quantization has not yet been approached in a manner that would retain the higher-dimensional properties of the worldvolume.

In this paper we do not quantize these new F-theories directly, but consider how uncompactified string amplitudes might be reproduced in a higher-dimensional worldvolume.  In particular, we want to see how at least some of the features of 2d conformal field theory might be applied.  The simplest way is to add equal numbers of bosonic and fermionic dimensions to the worldsheet in such a way that they cancel in quantum computations [7], in analogy to the way a similar procedure produced basic properties of string field theory when applied to the spacetime dimensions of lightcone string theory [8].  (When applied to ordinary field theory, this method applies to spin as well as coordinates, as a generalization of Feynman's original discovery of ghosts for Yang-Mills [9].)  This is a guess at how a ``covariant gauge" for F-theory might look, in contrast to the ``unitary gauge" (analogous to the lightcone gauge) where the worldvolume constraints are explicitly solved to eliminate the extra worldvolume dimensions.  Thus, although this generalization is in some sense trivial, it provides a plausible goal for a derivation based on applying any of the standard gauge-fixing procedures of quantum theory to F-theory.

Ü2. Worldvolume propagator

It's difficult to see how the standard closed-string amplitudes could be reproduced without the usual logarithmic propagator for $X$ on the worldsheet.  Adding equal (even) numbers of bosonic and fermionic dimensions to the worldsheet solves this in the standard Parisi-Sourlas way, as seen by defining the propagator by Fourier transformation of the inverse of the corresponding (massless) Klein-Gordon operator of the worldvolume:
$$ Çd^{d|d-2}p¼e^{ipÉz}{1\over p^2} = Ç_0^¥ d Çd^{d|d-2}p¼e^{ipÉz- p^2} ¾ -ln(z^2) $$
The ``d|d$-$2" refers to d bosonic and d$-$2 fermionic worldvolume dimensions, with ``$p^2$" (and $z^2$) using the OSp(d|d$-$2) metric.  (Of course, d=2 is the usual worldsheet.  We ignore questions of signature.  d is assumed to be even.)  Gaussian integration over $p$ gives a factor of $1/å $ to the power $str(I)$ = d$-$(d$-$2) = 2, yielding the same form for the result as for d=2 (where we dropped the usual divergent constant).

Ü3. Conformal invariance

As a generalization of the Sp(2)$^2$=SO(4) (again ignoring signature) M¬obius invariance of the worldsheet for the closed string, we look for OSp(d+2|d$-$2) ``conformal" invariance.  This will be used to fix the positions of 3 vertex operators: (1) The first is fixed by translations.  (2) Another is fixed by conformal boosts (as easily seen by considering inversions).  (3) For the third, we fix its norm with a scale transformation, and its direction with ``Lorentz".  This leaves a residual OSp(d$-$1|d$-$2) symmetry.

The usual closed string amplitudes all have products of holomorphic times antiholomorphic factors, giving $|z|^2$'s that easily translate into OSp squares $z^2$.  Things then work pretty much the same as for the usual conformal field theories in higher dimensions.  Of course, translation, scale, and Lorentz invariance are easy to check.  This leaves only conformal boosts; but invariance under them follows from invariance under inversions.  Then we have the usual
$$ z £ z' = {z\over z^2}âÜâz_{12}^2 £ {z_{12}^2\over z_1^2 z_2^2} $$
where $z_{12}­z_1-z_2$.  That leaves us with the measure, whose transformation gives a Jacobian
$$ d^{d|d-2}z £ d^{d|d-2}z' = d^{d|d-2}z¼sdet\left({»z'\over »z}\right) $$
where
$$ {»z'^a\over »z^b} = {1\over z^2}\left( ¶_b^a -2{z_b z^a\over z^2}\right) $$
and the $a,b$ indices run over d commuting and d$-$2 anticommuting values.
(There are the usual statistics ordering signs that we can handle by carefully watching indices.)  The superdeterminant of the first factor again gives the power $str(I)$, reproducing the d=2 result.  That of the second factor can be evaluated, e.g., by expanding in the second term, or choosing a specific direction for $z^a$:  Either way the result is $-1$ (again as for d=2).  After fixing the limits of integration, we then have the result resembling d=2,
$$ d^{d|d-2}z £ {d^{d|d-2}z\over (z^2)^2} $$
Invariance of amplitudes can be checked by performing transformations and checking cancelation of factors of $z_i^2$ under inversions.

Ü4. Amplitudes

As an example we evaluate the 4-point tachyon amplitude in the bosonic theory.  We start with the usual expression for the N-point amplitude, substituting just $|z|^2£z^2$:
$$ A_N = z_{1,N-1}^2 z_{N-1,N}^2 z_{N,1}^2Çd^{(N-3)(d|d-2)}zÞ_{i<j}(z_{ij}^2)^{Œ'k_iÉk_j/2} $$
where we have chosen $z_1,z_{N-1},z_N$ as the 3 fixed $z$'s.  Here $k_iÉk_j$ refers to the spacetime inner product of the usual momenta, which we don't discuss here.  (We assume covariant gauges have been chosen for the $X$ worldvolume gauge fields, so they appear with naive index contraction.)  The momentum-dependent factors follow from the logarithmic propagators produced by the usual vertex operators $e^{ikÉX(z)}$; the measure factor in front has been chosen for OSp(d+2|d$-$2) worldvolume conformal invariance.   This invariance works the same way as for d=2, since the transformations for $z_{ij}^2$ and $d^{d|d-2}z_i$ are the same.

For the case N=4, after taking $z_i£(0,z,1,¥)$, where ``1" means a unit vector in some fixed (bosonic) direction, this becomes the ``usual"
$$ A_4 = Çd^{d|d-2}zÊ(z^2)^{-üŒ(s)-1}[(1-z)^2]^{-üŒ(t)-1} $$
where
$$ Œ(s) = üŒ's + 2¼,âs+t+u = -{16\over Œ'} $$
$s=-(k_1+k_2)^2$, etc., and $Œ'$ would be the open-string slope.  

To evaluate we use the usual tricks:  Compare to a massless propagator correction in d|d$-$2 dimensions with ``1" the external momentum and $z$ the loop momentum, and the 2 internal propagators each loop-modified to some powers of momenta.  This suggests exponentiating the 2 factors with Schwinger parameters
$$ f^{-h} = {1\over ý(h)}Ç_0^¥ d Ê ^{h-1}e^{- f} $$
Again the worldvolume integral is Gaussian (now over $z$ instead of its conjugate $p$), so it again gives a d-independent result in terms of $str(I)=2$, the same value as for the worldsheet d=2.  (We only needed the fact that $z_3^2=1$, so choosing its direction was irrelevant.  But the same was true for d=2, where only $|z|^2=1$ for the ``2-vector" $z$ was needed.  More general fixed $z$'s require more complicated expressions for $A_4$, as follow from the more general expression given for $A_N$.)
The remaining integrals are thus the same.  (Combine the 2 Schwinger parameters into scaling parameter $Â$ and Feynman parameter $Œ$, $( _1, _2)=Â(Œ,1-Œ)$, etc.)  The final result is the usual Virasoro-Shapiro amplitude, identical to d=2.

Ü5. Conclusions

We have shown how F-theory (not just the exceptional supergravity sector) might be applied in a gauge where the worldvolume is not reduced to the worldsheet, by including ``zeroth-quantized" ghosts: not fermionic partners for second-quantized fields $Ä(x)$, nor for first-quantized worldvolume fields $X(z)$, but for the worldvolume coordinates $z$.  (The relevance of zeroth-quantization for strings [10] and higher-dimensional worldvolumes [11] was considered previously.)

But we need also at least the first-quantized ghosts $c(z)$ and $b(z)$ for deriving the measure factor ($z_{1,N-1}^2 z_{N-1,N}^2 z_{N,1}^2$ for trees, used above) and evaluating loops.  The naive generalization of the worldsheet ghosts seems noncovariant:  The 2d ``conformal gauge" is actually a temporal gauge (Gaussian normal coordinates for the scale-free metric).  In general dimensions it leads to a ghost kinetic term of the form
$$ b^{0a}[üstr(I)»_{(0}c_{a)} -ú_{0a}»^b c_b] $$
(For the usual worldsheet this can be rewritten in a manifestly covariant form.)  The ghosts $b$ and $c$ are each the same in number as the worldvolume coordinates.  In our case $str(I)=2$.  The equations for $c$ are weaker than the covariant ones that would correspond to fixing all the conformal metric,
$$ üstr(I)»_{(a}c_{b)} -ú_{ab}»^b c_b = 0 $$
These would be too strong, as they would restrict $c$ to have only the ``zero-modes" corresponding to the conformal group, and no oscillator modes.

An alternative is suggested by noting that in the 2d case these ghost equations
$$ »_0 c_0 + »_i c_i = 0 = »_0 c_i + »_i c_0 $$
can be rewritten, after replacing $c_0£-c_0$, as
$$ »^a c_a = 0 = »_{[0}c_{i]} $$
This suggests the slight generalization
$$ »^a c_a = 0 = »_{[a}c_{b]}âÜâc_a = »_a c¼,â»^a »_a c = 0 $$
This corresponds to the ghosts suggested in [8].  However, in d>2 this removes some of the conformal zero-modes, but only those of the residual OSp(d$-$1|d$-$2) (although the useful zero-modes appear in funny places, due to the noncovariance of the $c_0$ redefinition).

The extra zeroth-quantized ghosts for canceling extra worldvolume dimensions d in a covariant gauge suggests the existence of analogous extra first-quantized ghosts in a covariant gauge for the extra spacetime dimensions D.

ÜAcknowledgments

This work was supported in part by National Science Foundation Grant No.¼PHY-1316617.

\refs

£1 C. Vafa, \NP 469 (1996) 403, \xxxlink{hep-th/9602022};\\
C.M. Hull, \NP 468 (1996) 113, \xxxlink{hep-th/9512181};\\
M.P. Blencowe and M.J. Duff, \NP 310 (1988) 387.

£2 E. Cremmer and B. Julia, \PL 80B (1978) 48, \NP 159 (1979) 141;\\
B. Julia, Group disintegrations, in ÓNuffield Workshop on Superspace and SupergravityÕ, eds. S.W. Hawking and M. Rocek, Cambridge (Cambridge Univ., 1980) 331;\\
W. Siegel, \NP 177 (1981) 325;\\
P.C. West, ÓClass.Quant.Grav.Õ É18 (2001) 4443, \xxxlink{hep-th/0104081},\\ \PL 575B (2003) 333, \xxxlink{hep-th/0307098};\\
C.M. Hull, ÓJHEPÕ É0707 (2007) 079, \xxxlink{hep-th/0701203};\\
D.S. Berman and M.J. Perry, ÓJHEPÕ É1106 (2011) 074, \xxxlink{1008.1763} [hep-th];\\
A. Coimbra, C. Strickland-Constable, and D. Waldram, ÓJHEPÕ É1402 (2014) 054,\\ \xxxlink{1112.3989} [hep-th];\\
D.S. Berman, M. Cederwall, A. Kleinschmidt, and D.C. Thompson, ÓJHEPÕ É1301 (2013) 064, \xxxlink{1208.5884} [hep-th];\\
O. Hohm and H. Samtleben, \PR 111 (2013) 231601, \xxxlink{1308.1673} [hep-th], \PRD 89 (2014) 066016, \xxxlink{1312.0614} [hep-th],
\PRD 89 (2014) 066017, \xxxlink{1312.4542} [hep-th], \PRD 90 (2014) 066002, \xxxlink{1406.3348} [hep-th];\\
H. Godazgar, M. Godazgar, O. Hohm, H. Nicolai, and H. Samtleben, ÓJHEPÕ É1409 (2014) 044, \xxxlink{1406.3235} [hep-th];\\
E. Musaev and H. Samtleben, ÓJHEPÕ É1503 (2015) 027, \xxxlink{1412.7286} [hep-th].

£3 P.A.M. Dirac, ÓProc.Roy.Soc.Lond.Õ ÉA268 (1962) 57;\\
E. Bergshoeff, E. Sezgin, and P.K. Townsend, \PL B189 (1987) 75.

£4 B. de Wit, J. Hoppe, and H. Nicolai, \NP 305 (1988) 545;\\
Luca Mezin'cescu, Rafael I. Nepomechie, and P. van Nieuwenhuizen, \NP 309 (1988) 317.

£5 W.D.Linch, III and W. Siegel, \xxxlink{1501.02761} [hep-th], \xxxlink{1502.00510} [hep-th], \xxxlink{1503.00940} [hep-th], \xxxlink{1507.01669} [hep-th]. 

£6 W. Siegel, \PRD 47 (1993) 5453, \xxxlink{hep-th/9302036}; \PRD 48 (1993) 2826, \xxxlink{hep-th/9305073};
Manifest duality in low-energy superstrings, in ÓStrings '93Õ, eds. M.B. Halpern, G. Rivlis, and A. Sevrin, Berkeley (World Scientific, 1995) 353, \xxxlink{hep-th/9308133}.

£7 G. Parisi and N. Sourlas, \PR 43 (1979) 744.

£8 W. Siegel, \PL 142B (1984) 276.

£9 R.P. Feynman, ÓActa Phys. Polon.Õ É24 (1963) 697.

£10 M. Green, \NP 293 (1987) 593.

£11 D. Kutasov and E.J. Martinec, \NP 477 (1996) 652, \xxxlink{hep-th/9602049}, ÓClass.Quant.Grav.Õ É14 (1997) 2483, \xxxlink{hep-th/9612102};\\
D. Kutasov, E.J. Martinec, and M. O'Loughlin, \NP 477 (1996) 675,\\ \xxxlink{hep-th/9603116}.

\bye